\documentclass[twocolumn,aps,prc,preprintnumbers]{revtex4}%
\usepackage{amssymb}
\usepackage{amsmath}
\usepackage{amsfonts}
\usepackage{graphicx}
\usepackage{dcolumn}
\usepackage{bm}%
\providecommand{\U}[1]{\protect\rule{.1in}{.1in}}
\providecommand{\U}[1]{\protect\rule{.1in}{.1in}} \textheight 9.7in
\begin{document}
\title{Fusion11 Conference Summary}
\author{Carlos A.~Bertulani}
\email{carlos_bertulani@tamu-commerce.edu}
\address{Department of Physics and Astronomy, Texas A\&M University-Commerce, Commerce,
TX 75429, USA}

\begin{abstract}
A summary account of the conference {\it ``Fusion11"}, held in  Saint Malo, France, May 2-6, 2011.

\end{abstract}

\maketitle

\section{Introduction}
\label{intro}
\label{sec:i}
{\sl Navin Alahari} (Chair) made a short overview of what the community expects to learn, what are the main experimental and theoretical difficulties, and how we can proceed in the future. In particular, he pointed out that he organized this meeting because ``{\it there are some questions that cannot be answered with Google}". Reluctantly, I tend to agree. 

There were 77 talks by experts in the field. I will comment on a few of the physics topics discussed during the presentations. My description is evidently limited in scope due to the short space.
\section{New facilities}
\label{sec:1}
{\sl Sidney Gales} presented an overview of the future of the SPIRAL  2 facility at GANIL, France, which promises to break the  high intensity frontier for both stable and  ``{\it exotic}" beams. The construction will occur in two phases, with the production of radioactive beams planned for 2015. Physics with radioactive and stable-ion high-intensity  beams at 1-20 MeV/nucleon will allow, among many other processes, the study of fusion-evaporation with very small cross sections, possibly at the nb level. 

{\sl Peter Thirolf} introduced the Extreme Light Infrastructure (ELI-Nuclear Physics) facility to be built in Bucharest, Romania, by the European Community.  This facility will be devoted to research with high-intensity lasers (20 petawatts), up to 5 orders of magnitude higher than today's laser intensity. It would eventually allow to study laser induced fission-fusion events. If all goes as planned, the facility would be operational by 2015.

{\sl P. Monier-Garbet} also described the outcome of a new heavy-weight facility: ITER, the {\it international project for thermonuclear fusion}. This project (presently estimated at 15 billion euros) will open a new epoch for studies of fusion plasmas. ITER is designed to confine a deuteron-tritium plasma in which $\alpha$-particle heating dominates all other forms of plasma heating. ITER's principal goal is to design, construct and operate a tokamak
experiment at a scale which satisfies the objective  of using {\it fusion energy for peaceful purposes}. ITER's basic research will be atomic physics and material science. The generation of commercial energy by using thermonuclear fusion will be a quest for future post-ITER projects.  Humanity will profit enormously if this is realized.

\section{Fusion Cross Sections}
\label{sec:2}
As mentioned by {\sl S. Umar}, no practical ab-initio many-body theory for fusion exists. All approaches involve two prongs: a) Calculate an ion-ion (usually one-dimensional) phenomenological potential (Wood-Saxon, proximity, folding, Bass, etc.) using frozen densities, or microscopic, macroscopic-microscopic methods using collective variables (CHF, ATDHF, empirical methods), and b) Employ quantum mechanical tunneling methods for the reduced one-body problem (WKB, IWBC),  incorporating quantum mechanical processes by hand, including neutron transfer and excitations of the entrance channel nuclei (CC).\subsection{Barrier Penetration Model (BPM)}
\label{sec:2.1}

Fusion cross sections can be calculated from the equation
\begin{equation}
\sigma_F(E)=\pi {\lambda}^2 \sum_\ell (2\ell +1) P_\ell (E) , \label{eqnf}
\end{equation}
where $E$ is the center of mass energy, $\lambda=\sqrt{\hbar^2/2mE}$ is the reduced wavelength and $\ell=0,1,2,\cdots$.
The cross section is proportional to $\pi\lambda^2$, the area of the quantum wave. Each part of the wave corresponds to different impact parameters having different probabilities for fusion. As the impact parameter increases, so does the angular momentum, hence the reason for the $2\ell+1$ term. $P_\ell(E)$ is the probability that  fusion occurs at a given impact parameter, or angular momentum. The barrier penetration method (BPM) assumes that fusion occurs when the particle (with mass $m$) penetrates the Coulomb barrier and $P_\ell$ is calculated in a one-dimensional potential model, e.g. by using the WKB approximation or alike. 

\begin{figure}
\begin{center}
\resizebox{0.6\columnwidth}{!}{%
\includegraphics{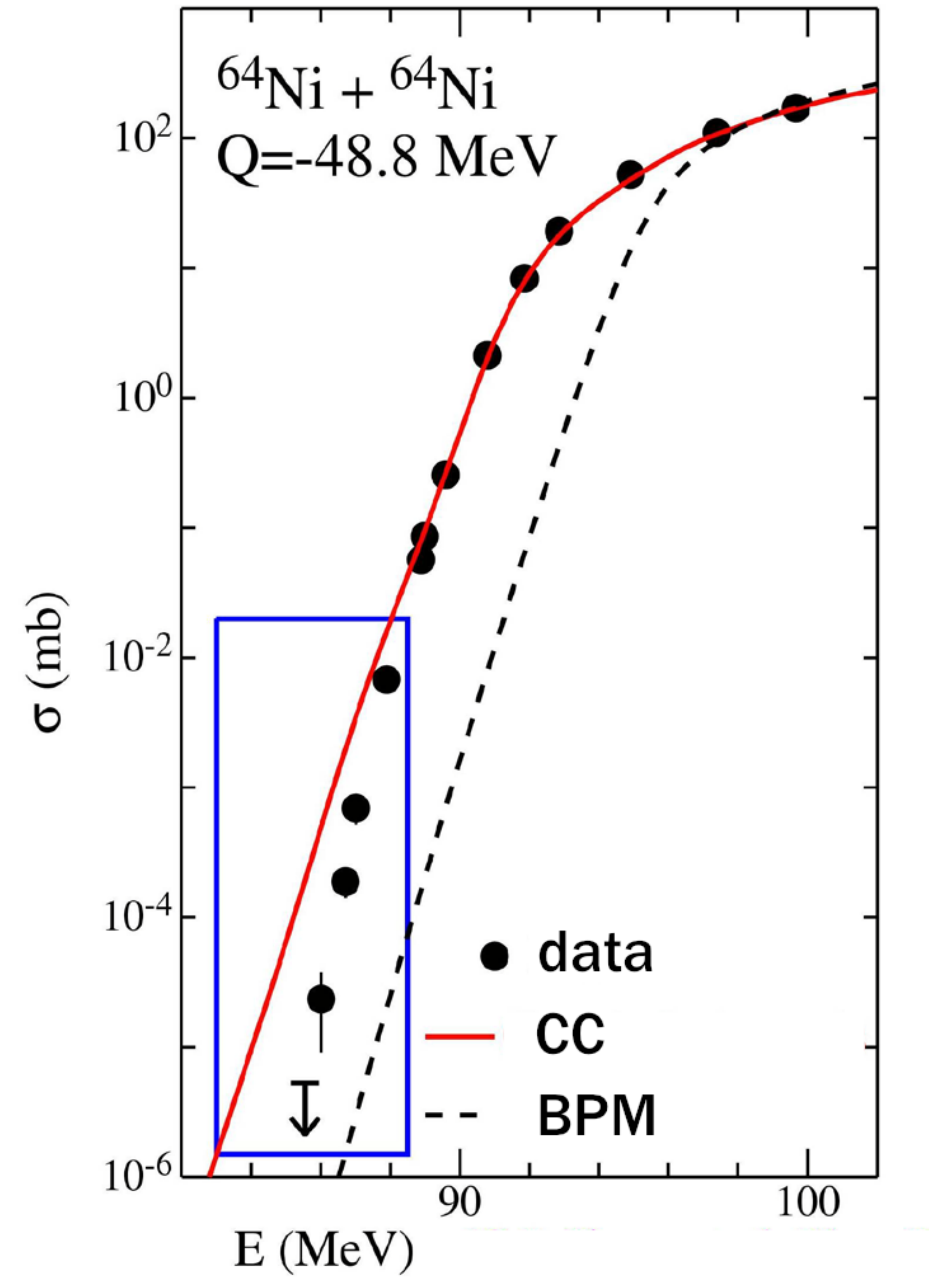}}
\end{center}
\caption{Fusion cross section of $^{64}$+$^{64}$Ni as a function of the center of mass energy \cite{Ji04}. The dashed (solid) curve is a BPM (coupled-channel) calculation (Courtesy of {\sl C.L. Jiang}). }
\label{fig:1}       
\end{figure}

From $\sigma_\ell =\pi {\lambda}^2 (2\ell+1) P_\ell$ one can calculate the average value of $\ell$ from $\langle \ell(E)\rangle = \sum_\ell \ell \sigma_\ell/\sum \sigma_\ell$ and many other relevant quantities.  Sometimes, for a better visualization, or for extrapolation to low energies, one uses the concept of {\it astrophysical S-factor}, redefining the cross section as   
\begin{equation}
\sigma_F(E)={1\over E} S(E)\exp\left[ -2\pi \eta(E)\right], \label{SE}
\end{equation} 
where $\eta(E)=Z_1Z_2e^2/\hbar v$, with $v$ being the relative velocity. The exponential function is an approximation to $P_0(E)$ for a square-well nuclear potential plus Coulomb potential, whereas the factor $1/E$ is proportional to the area appearing in Eq. \ref{eqnf}.   Sometimes one just plots the cross section as $L(E)=\ln [E\sigma_F(E)]$.

\subsection{Optical Potentials (OP)}
\label{sec:2.2}
In order to use Eq. \ref{eqnf} one needs the nucleus-nucleus potential. This is a {\it badly known beast}. In some cases, one includes the effects of other relevant non-fusion channels (which might be a very large number, say $\infty$). Then one adds an imaginary part to the real potential and it becomes much more than a beast; something like a combination of the {\it Devil$^{\small \copyright}$, the Alien$^{\small \copyright}$} and {\it the Predator$^{\small \copyright}$}.  Some have tried to tame this thing from first principles. But, except for few heroic attempts, we seem to have given up. We just  fit whatever we can fit and we get whatever parameters of a potential function  we can. Then we wisely call it the ``OP". 

In this meeting there were some discussions on the OP appropriate for fusion reactions. {\sl D. Pereira} reported  extensions of the S\~ao Paulo potential, and a  new approach for its imaginary part to account for surface dissipative processes in heavy ion reactions. {\sl Pereira} showed that this method could reproduce the $^7$Li(25 MeV)+$^{120}$Sn and the $^4$He(230 MeV)+$^{12}$C elastic scattering data rather well.  The  {\it S\~ao Paulo potential} has become popular in fusion reactions, due to its simple form \cite{SPP}. So, does the M3Y potential, which was originally developed to treat heavy ion collisions at $E\sim 100$ MeV/nucleon. It does not really matter: {\it we love delta-function} potentials. In fact, looking from very far away, the nucleon-nucleon potential does  indeed resembles a delta-function. An that is why we use them in effective field theories, mean field calculations (Skyrrme interaction), and in many other areas. Simple.

{\sl G. Marti} presented results for $^{8,7}$Li +$^{20}$Se elastic scattering at low energies. The data seem to show another effect which goes by the name ``{\it threshold anomaly}". This was also discussed in the talk by {\sl M. Sinha}. After fitting the OP to elastic scattering data, the anomaly appears in plots of energy dependence of the real, $V$, and imaginary, $W$, separate parts of the OP at a certain nucleus-nucleus distance. As the bombarding  energy decreases, $W$ remains rather flat and  below a certain energy value it quickly drops to zero, whereas $V$ increases slowly and below the same energy reverses its trend \cite{Na85}. The behavior of $V$ and $W$ is compatible with dispersion relations (Yes. We believe that the {\it OP is an analytic function} in a complex plane!). 

Fusion with loosely bound nuclei shows a ``{\it breakup threshold anomaly}", meaning a small rise of the potentials before the anomaly threshold \cite{Hu06}. I am not sure what kind of physics one wants to extract from (breakup) threshold anomalies. The physics seems to be non-universal, e.g., {\sl Marti} claims that, for the elastic scattering of $^7$Li the behavior of both types of potentials as a function of energy is compatible with the presence of the threshold anomaly. But {\sl C.S. Palshetkar} has reported the absence of the anomaly for $^9$Be+$^{89}$Y system. Do you want to know what anomaly means? ``Look! I have found something new!".

\begin{figure}
\begin{center}
\resizebox{0.8\columnwidth}{!}{%
\includegraphics{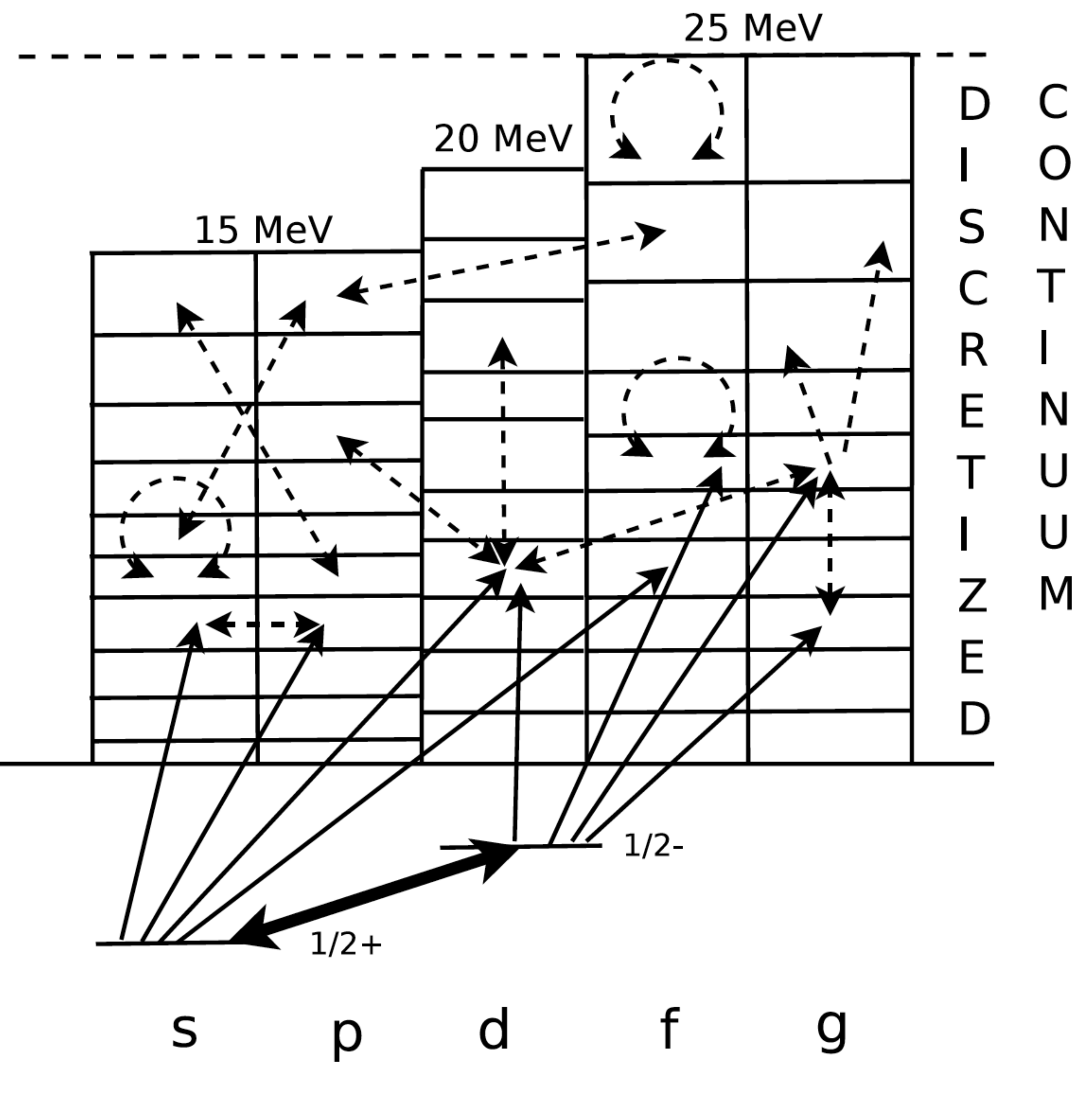}}
\end{center}
\caption{Schematic representation of CDCC transitions. (Courtesy of {\sl F. Canto}).}
\label{fig:1}       
\end{figure}

{\sl C.J. Lin} also explored the concept of {\it ``surface anomaly"}. What is anomalous here? The diffuseness parameters $a$ of interaction potentials extracted from the fusion data at not only high energies but also extremely low energies are prominent larger than the traditional value (0.65 fm).

{\sl S. Umar} discussed results from Time-Dependent Hartree-Fock (TDHF) calculations using Skyrme functionals. His goal was to deduce microscopically the  potential barriers needed to calculate fusion reaction cross sections. Although heavy systems pose a greater challenge, such microscopic calculations may provide an insight into these collisions.

\subsection{Channel Coupling (CC)}
 \label{sec:2.3}
It is well know that the BPM does not work. A good example is shown in figure \ref{fig:1}, taken from Ref. \cite{Ji04}. Only by including coupling to other channels, the fusion cross sections can be reproduced. In fact, sub-barrier fusion enhancement can be easily understood. Assume that the initial channel energy $E$  splits into two channel energies  during fusion, $E_1=E+\Delta$ and $E_2=E-\Delta$. Then the cross section becomes an approximate energy average of the BPM, $\sigma_F(E)\sim[\sigma_F(E_1)+\sigma_F(E_2)]/2$. Due to the exponential behavior of tunneling probabilities, the decreased cross section $\sigma_F(E_1)<\sigma_F(E)$ more than compensates the increase $\sigma_F(E_2)>\sigma_F(E)$ due to the reduction of the barrier height.

\begin{figure}
\resizebox{1.1\columnwidth}{!}{%
\includegraphics{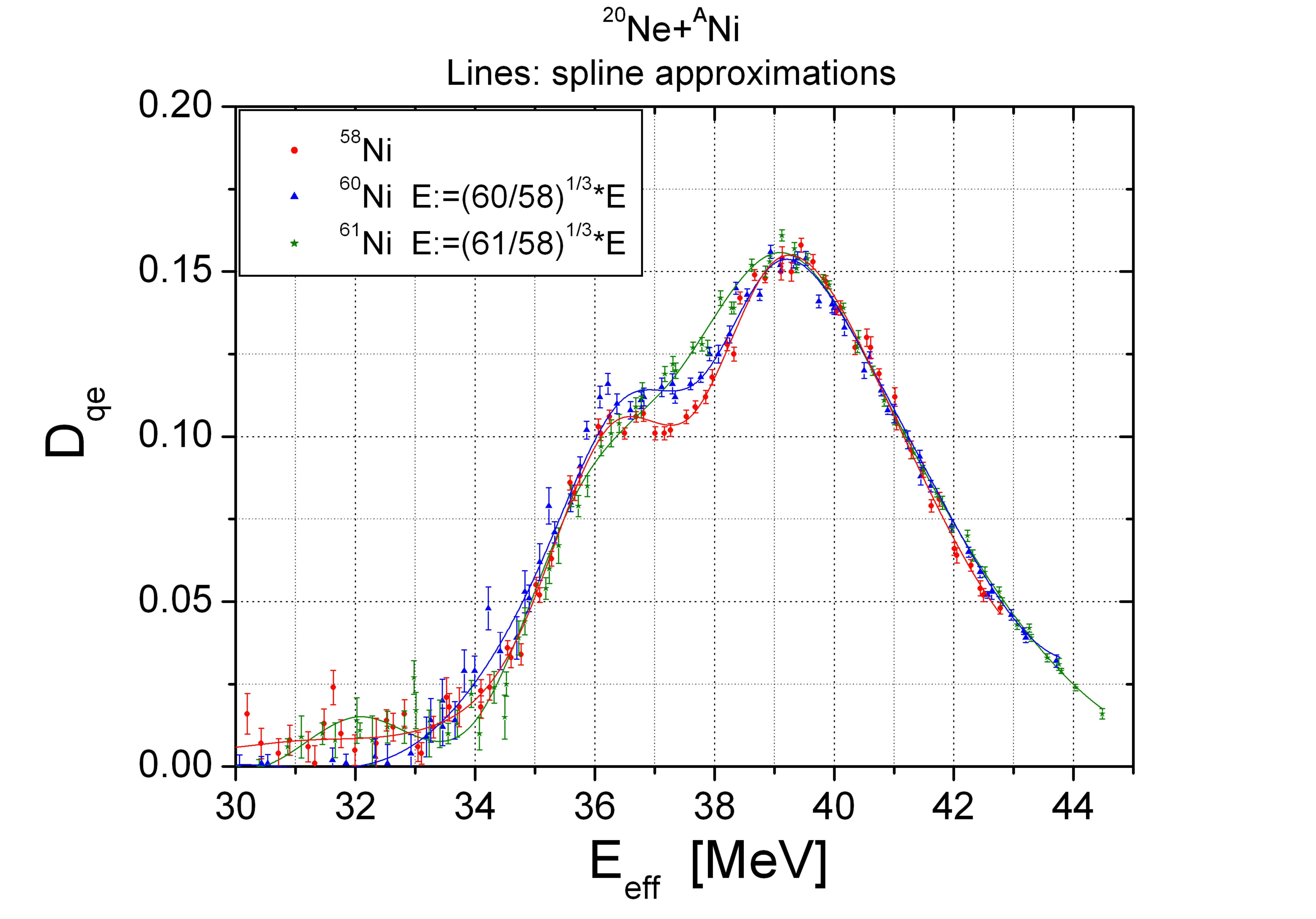}}
\caption{Barrier distribution function from  $^{20}$Ne+$^{58,60,61}$Ni quasi-elastic scattering measurements (Courtesy of {\sl A. Trczci\'nska}).}
\label{fig:bd}       
\end{figure}

In  coupled channels schemes  one expands the total wavefunction for the system as
\begin{equation}
\Psi=\sum_{i,k} a_i(\alpha, q_k)\phi(\alpha, q_k),
\end{equation} 
where $\phi$ form the channel  basis, $\alpha$ is a dynamical variable (e.g., the distance between the nuclei), and $q_k$ are intrinsic coordinates. Inserting this expansion in the Schr\"odinger equation yields  a set of CC equations in the form
\begin{equation}
{da_k\over d\alpha}=\sum_j a_j  \ \langle \phi_k \left| U \right| \phi_j \rangle \ e^{E_\alpha \alpha}, \label{cc}
\end{equation}
where $U$ is whatever potential couples the channels $k$ and $j$ and $E_\alpha=E_\alpha^{(k)}-E_\alpha^{(j)}$ is some sort of transition energy, or transition momentum. In the presence of continuum states, continuum-continuum coupling(relevant for breakup channels) can be included by {\it discretizing the continuum}. This goes by the name of Continuum Discretized Coupled-Channels (CDCC) calculations.  Fancy, but also vulgar.
There are several variations of CC equations, e.g., a set of differential equations for the wavefunctions, instead of using basis amplitudes.

Coupled channels calculations were discussed in details by {\sl L.F. Canto} in his review talk  of fusion reactions.  CDCC calculations for d+A with rotational and vibrational channels was reported by {\sl P. ChauHuu-Tai}, while {\sl A. Moro} discussed a simultaneous analysis of elastic, breakup, and fusion channels for the $^6$He+$^{208}$Pb reaction at energies near the Coulomb barrier. It seems that continuum-continuum couplings hinder fusion,  specially above the barrier, but the reason is not well understood.

Coupled channels calculations with a large number of channels in continuum couplings, is one of the least controllable calculations in physics. Anything can happen because of the phases of matrix elements: the couplings can add as $+-+--++-+$ (destructive) or as $++++-++++$ or as $----+----$ (constructive), depending on the system and on the nuclear model. Such suppressions or enhancements are difficult to understand. {\sl It is simply disgusting}. Maybe we should not try to understand. But as Wigner once said: {\it ``it is nice that your computer can understand this stuff. But I wish I could understand it myself"}.

CDCC calculations were also reported by {\sl K. Hagino} for $^{16}$O+$^{208}$Pb. He included collective and  non-collective states, weakly coupled, a total of 64 non-collective levels up to 7 MeV, with a nearly ``complete" level scheme both for the excitation energies $E^*$ as for the  $\beta_\lambda$  parameters  of a deformed nuclear model. He concluded that the energy dependence of fusion cross section is not altered much by these couplings.

Coupled channels calculations were also reported by {\sl T. Ichikawa}. {\sl C.J. Lin} also claimed that the ``surface anomaly" (see section \ref{sec:2.2}) disappears when CC are taken into account.

\subsection{Barrier Distribution}
 \label{sec:2.4}
One can try to extract some extra juice from a set of experimental data by recasting them in a different way. For example, assume an extreme classical model in which fusion occurs. Also assume that fusion is hindered by Coulomb recoil. Then the fusion cross section is given approximately by $\sigma_F \sim (1- V_B/E)\Theta(E-V_B)$ where $\Theta(x)=0$ for $x\le 0$, and $\Theta=1$, otherwise. $V_B=Z_1Z_2e^2/R$ is the Coulomb barrier height and $R$ is the touching distance between the nuclei. In this approximation, the second derivative of $\sigma_F$ yields  $${d^2\sigma_F\over dE^2}\sim \delta(E-V_B),$$ i.e. it spikes at the value of the Coulomb barrier height. Quantum mechanics smears out the $\delta$-function \cite{Es78}, but $d^2\sigma_F/dE^2$ should still peak at the position of the Coulomb barrier height. Coupling to other channels fragments the peak and broadens it further, as seen in experiments.  $d^2\sigma_F/dE^2$ is thus a probe of {\it barrier heights} \cite{RS91}.

\begin{figure}
\resizebox{1.0\columnwidth}{!}{%
\includegraphics{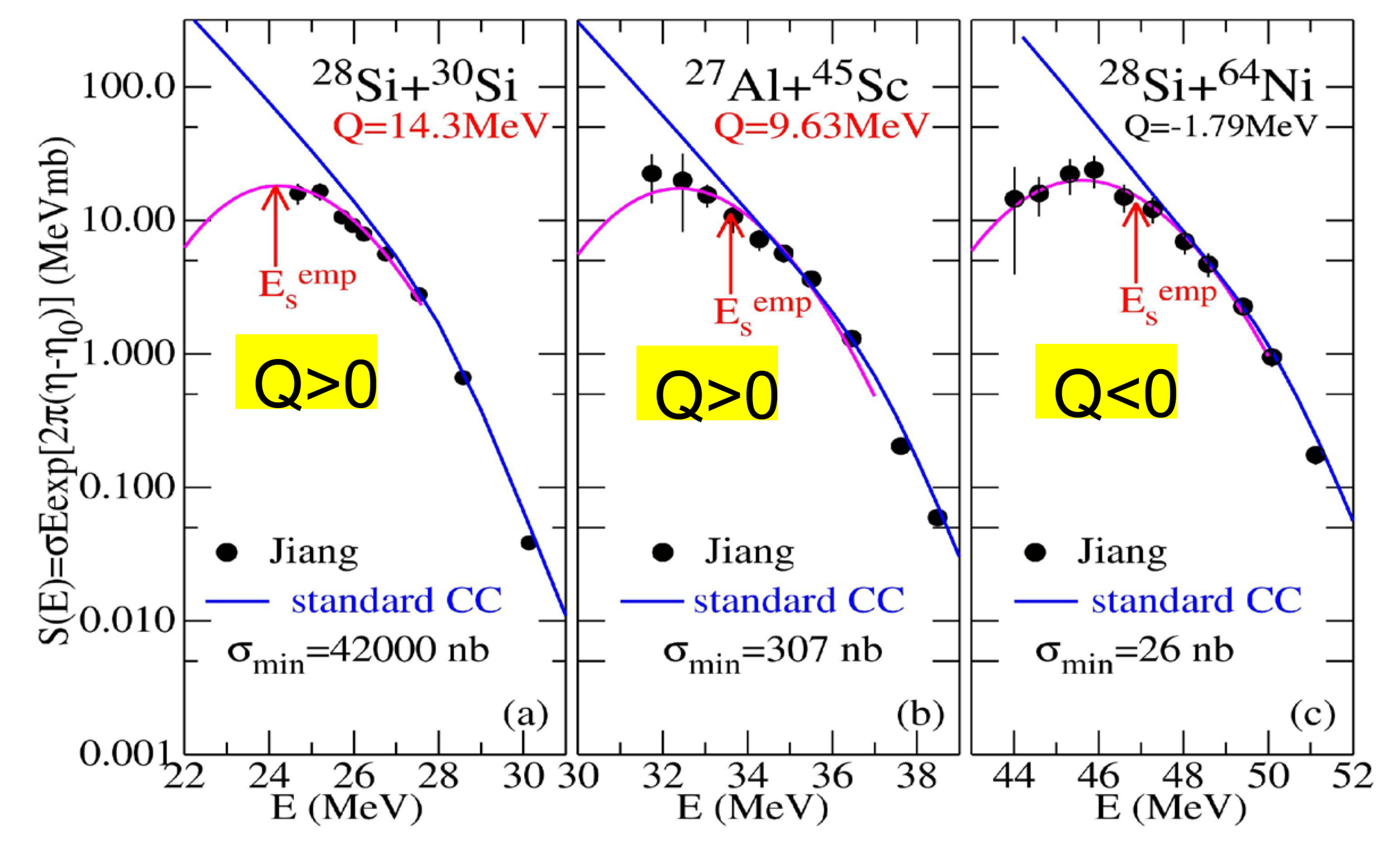}}
\caption{$S$-factors for $^{28}$Si+$^{30}$Si, $^{27}$Al+$^{45}$Sc and $^{28}$Si+$^{64}$Ni. (Courtesy of {\sl C.L. Jiang}).}
\label{fig:ji}       
\end{figure}

{\it Quasi-elastic scattering} maybe defined as a sum of all reaction processes other than fusion, e.g. elastic scattering, inelastic scattering, transfer, breakup, and so on. Thus fusion and quasi-elastic (qe) scattering are complementary to each other: if $d^2\sigma_F/dE^2\sim \delta(E-V_B)$, a similar relationship should be applicable to $\sigma_{qe}$. It turns out that this is true for the ratio of $\sigma_{qe}$ and $\sigma_{Ruth}(E)$. This reasoning leads to 
\begin{equation}
{d\over dE}\left[{\sigma_{qe}(E)\over \sigma_{Ruth}(E)}\right] \sim \delta(E-V_B).
\end{equation} 
The advantage of using $D_{eq}=-d[\sigma_{qe}/\sigma_{Ruth}]/dE$ over $d^2\sigma_F/dE^2$ is that less data accuracy is required: a first derivative is much easier to do. $\sigma_{qe}$ is also a sum of everything: a very simple charged-particle detector can measure it \cite{Tim95}.  Measuring $\sigma_F$ requires a specialized recoil separator to separate evaporation residues (ER) from the incident beam: ER  and fission for heavy systems.

Data for barrier distribution were presented by {\sl G. Montagnoli} for $^{40}$Ca+$^{40}$Ca, by {\sl C.J. Lin} for $^{6,7}$Li+$^{208}$Pb, $^{209}$Bi and by {\sl A. Trczci\'nska} for $^{20}$Ne+$^{58,60,61}$Ni (see figure \ref{fig:bd}). Remarkable agreement between  $d[\sigma_{qe}/\sigma_R]/dE$ and $d^2\sigma_F/dE^2$ is seen in most data. {\sl Montagnoli} infers from her data that below the barrier, the log slope of the excitation function for $^{40}$Ca+$^{40}$Ca has a small plateau and starts again increasing at lower energies. 
{\sl A. Trczci\'nska}  concludes that for $^{58}$Ni the distribution has a clearly visible ``structure" (figure \ref{fig:bd}),
whereas for heavier Ni isotopes the structure is smoothed out (for $^{60}$Ni  partly, for $^{61}$Ni  completely).
 
\begin{figure}
\begin{center}\resizebox{0.9\columnwidth}{!}{%
\includegraphics{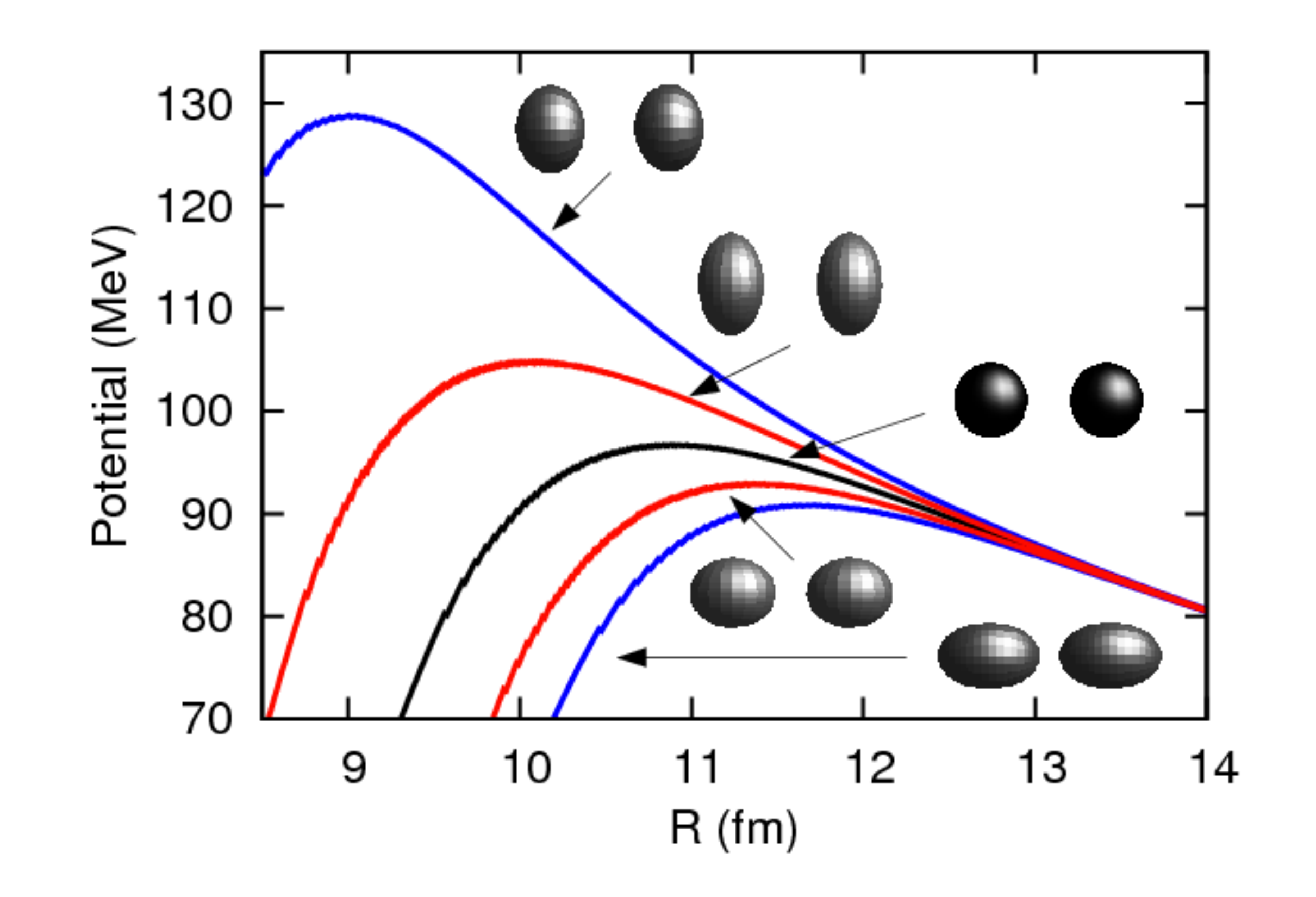}}
\end{center}\caption{Importance of deformation and orientation on the relative nucleus-nucleus potential. (Courtesy of {\sl B. Yilmaz}).}
\label{fig:or}       
\end{figure}

\subsection{Fusion Hindrance}
 \label{sec:2.5}
In figure \ref{fig:1} one notices that CC calculations do not quite reproduce the data, specially at deep sub-barrier energies. Our colleagues are working hard to gather data and understand this phenomenon. Cross sections need to be measured at very low energies, a real challenge for accelerator and detector development. As for theorists, there seems to be two main streams of thought. Well, just one: do everything as usual, but modify what happens when nuclei come very close, below the touching distance. A popular story goes like this:  build a potential from the M3Y interaction (there is a huge {\it M3Y fan-club} out there). Then claim that Pauli principle (quoting {\sl C. Simenel}: {\it if you do not know what to ask, ask me about the Pauli principle}) modifies the interior of the potential, decreasing its depth. This solves something, as M3Y yields a too deep central part of the potential. In Ref. \cite{EB07} it was also shown that this procedure also reproduces the nuclear incompressibility and the fusion hindrance in $^{16}$O+$^{208}$Pb deep sub-barrier fusion. 

In Ref. \cite{IHI07} the way to explain fusion hindrance was to assume a neck formation and the onset of damping for distances smaller than the touching radius. In summary, below the touching distance, something weird ({\it damping is a sort of physics hidden under a knob}) happens, leading to fusion hindrance.

{\sl C.L. Jiang} claims that all $Q<0$ fusion systems  must have an $S$ factor maximum, because $S(E)= 0$,  when $E=-Q$. Makes sense. This is shown in the rightmost panel of figure \ref{fig:ji}. As for $Q>0$ (most reactions in stars occur in this way) there is no such requirement. Usually the S-factors remain finite at $E= 0$. But as shown in the two left panels of figure \ref{fig:ji}, this is not really the case: a proof that fusion hindrance also occur for $Q>0$ fusion reactions. Intriguing. He also reported a first evidence of an S factor maximum in  the positive Q-value system $^{40}$Ca+$^{48}$Ca. But no firm conclusion can be made about the nature of the $S$ factor maximum, which is limited by background levels. {\sl A. Shrivastava} reported that  no hindrance was observed in the fusion of  $^7$Li($^{12}$C)+$^{197}$Pt. 

\begin{figure}
\resizebox{0.9\columnwidth}{!}{%
\includegraphics{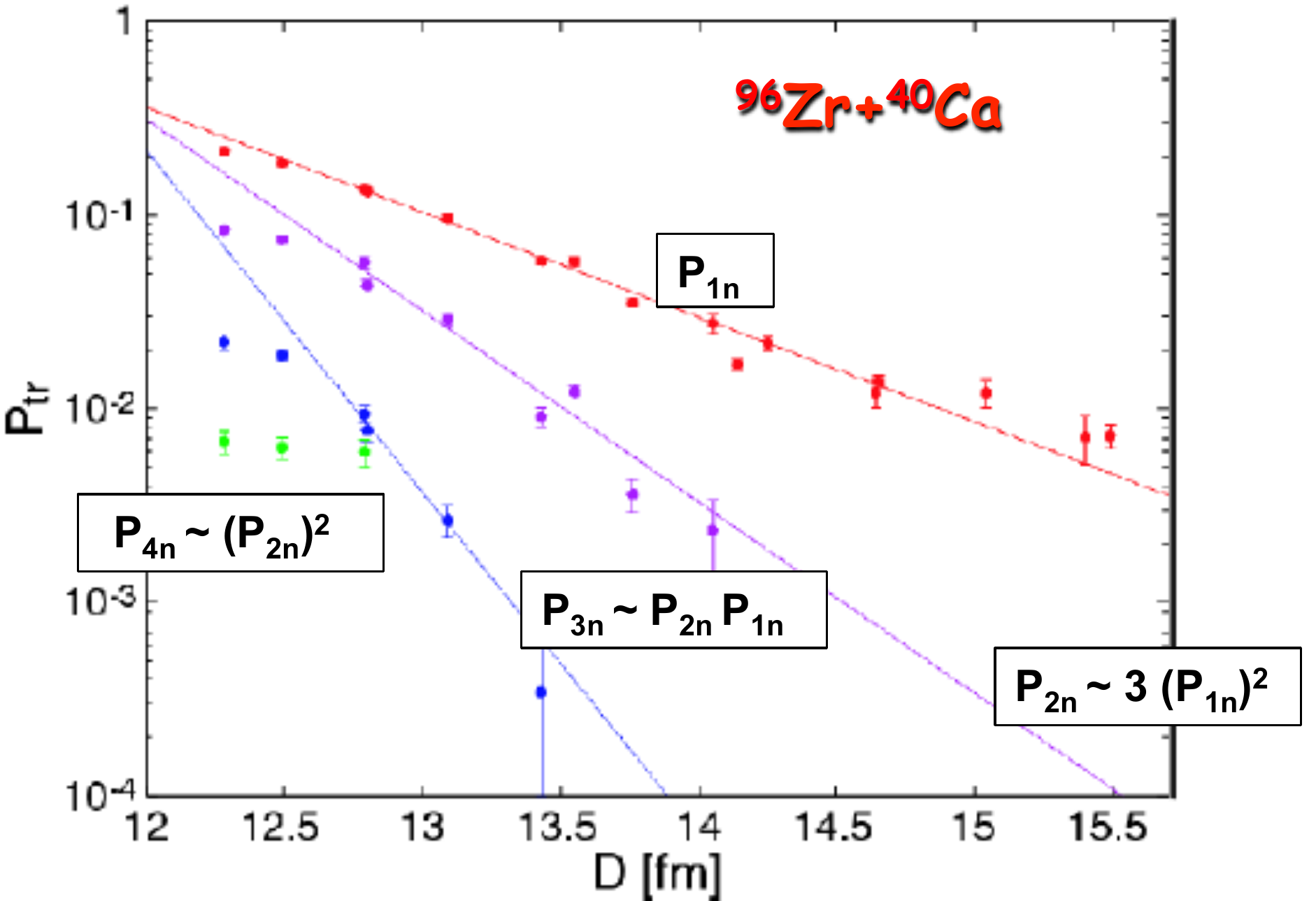}}
\caption{Transfer probabilities for multineutron transfer in $^{96}$Zr+$^{40}$Ca. (Courtesy of {\sl L. Corradi}).
}
\label{fig:co}       
\end{figure}

{\sl G.G. Adamian} discussed theoretical methods to calculate sub-barrier capture using  the {\it quantum diffusion approach}. This method takes into
account the fluctuation and dissipation effects in collisions which model the coupling of the relative motion with various channels. He claims that the effect of the change of fall rate of sub-barrier capture cross section should be in the data if we assume that the friction starts to act only when the colliding nuclei approach the barrier. But at extreme sub-barrier energies the experimental data still have large uncertainties to make a firm experimental conclusion about this effect \cite{SA10}.

{\sl B. Yilmaz} claimed that a classical treatment of the relative motion is a good approximation at near-barrier energies. Also, the quantal effects of the surface modes cannot be neglected. But it is possible to include them within a semi-classical approximation. In this model, the force on the nuclei follow the equation
\begin{eqnarray}
{dP\over dt}&=&-{dV_C(R)\over dR}-{dV_N(R,\Omega,\alpha_\lambda)\over dR}+{\ell(\ell+1)\hbar^2\over \mu R^3}
\nonumber \\ &-&\beta(R)P + F(t),
\end{eqnarray}
where $\Omega$ is the nuclear orientation, $\beta$ is a friction coefficient and $F$ is a fluctuating force satisfying $\langle F(t)\rangle=0$ and $\langle F(t) F(t)\rangle=2\mu \beta kT \delta(t-t') $, where the temperature $T$ is defined by the excitation energy. Quantum fluctuations are allowed through the coupling to surface mode amplitudes $\alpha_\lambda$. {\sl Yilmaz} showed the importance of deformation and orientation on the relative nucleus-nucleus potential, seen in figure \ref{fig:or}. This model is also able to explain fusion hindrance.
 
{\sl G.Montagnoli} reported that $^{48}$Ca+$^{48}$Ca new data are nicely reproduced with CC calculations using a shallow ion-ion potential, \`a la Ref. \cite{EB07}.   {\sl K. Washiyama} discussed a beyond mean-field approach  to heavy-ion reactions  around the Coulomb barrier with a stochastic description of energy dissipation. He obtains a dynamical reduction in the nucleus-nucleus potential, in good agreement with fusion experiments and also with mass variances $\sigma_{AA}^2$. {\sl D. Boilley}  showed that the neck fomation is a key ingredient and that the appearance of fusion hindrance sets some constraints on the fusion barriers.

\subsection{Transfer Channels}
 \label{sec:2.6}
 Assuming that $\alpha$ in Eq.  \ref{cc} is simply the time $t$, and using the first-Born approximation (i.e, taking $a_k\sim a_0\delta_{k0}$), the amplitude to excite the channel $\phi_k$ from an initial channel $\phi_0$ is given by $a_k=-i\hbar \int \langle \phi_0 |U|\phi_k\rangle \exp[i(E_k-E_0)t/\hbar]$. The Born approximation can be applied to transfer reactions, as shown by {\sl G. Pollarolo}. The probability to transfer a nucleon in nucleus A from channel $\alpha$ to a nucleon in nucleus B  in channel $\beta$ is given by
 \begin{equation}
P_{\beta\alpha}\sim \left| -i\hbar \int_{-\infty}^\infty dt F_{\beta\alpha}({\bf R}) \exp\left[i{E_\beta-E_\alpha)t\over \hbar}+(\cdots)\right]\right|^2, \label{transf}
 \end{equation} 
 where ${\bf R}$ is the nucleus-nucleus distance and $F_{\beta\alpha}({\bf R})$ is the from factor given by
  \begin{equation}
F_{\beta\alpha}({\bf R})=  \int d^3r e^{i{\bf Q}\cdot {\bf r}} \phi_\beta({\bf R}+{\bf r})\left( V_1-\langle U\rangle \right) \phi_\alpha({\bf r}),  \label{transf2}
\end{equation} 
where ${\bf Q}$ is the momentum transfer in the reaction, $U$ is the total (optical) potential, and $V_1$ is the potential of the nucleon with one of the nuclei. Why not $V_2$? In the literature, using $V_1$ ($V_2$) goes by the name {\it ``prior"(``post)-form}. It has been shown in the past that the post and prior forms of breakup and transfer reactions lead to the same result.  You should not worry with this. But you should worry with the $(\cdots)$ in Eq. \ref{transf}. They  are often associated with an {\it Uncontrolled Theoretical Ignorance} (UTI). Dangerous stuff.

\begin{figure}
\resizebox{1.0\columnwidth}{!}{%
\includegraphics{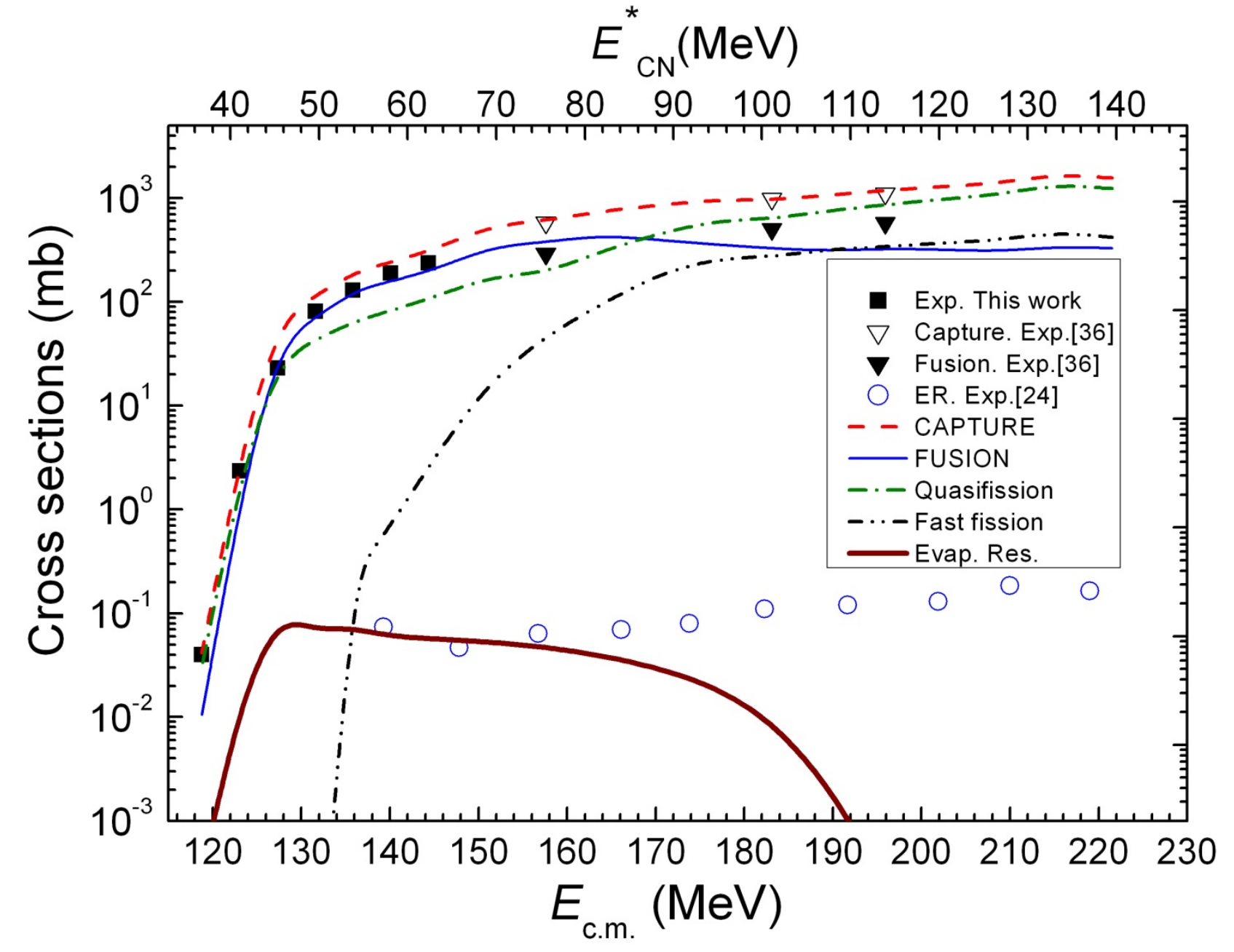}}
\caption{Interpretation of the measured capture and fusion excitation functions by description of evaporation residue cross sections. (Courtesy of {\sl H. Zhang}).
}
\label{fig:zh}       
\end{figure}

In figure \ref{fig:co}, shown by {\sl L. Corradi}, one sees the probabilities for multinucleon transfer in $^{96}$Zr+$^{40}$Ca, as a function of the closest approach distance $D=(Z_1Z_2e^2/2E)  [1+1/\sin (\theta/2)]$. Transfer is most likely to occur when the nuclei are at their closest point, $D$. The tunneling probability depends exponentially on this distance, $P_{tr}/\sin(\theta/2)\sim \exp(-2\alpha D)$. This approximation arises from Eqs. \ref{transf} and \ref{transf2}. If one neglects correlations, two-nucleon transfer probabilities are given in terms one-nucleon transfer probabilities: $P_{2n} = \left( P_{1n}\right)^2$. For three-nucleon transfer $P_{3n} = P_{1n}P_{2n}$, and so on. These are shown by the straight lines in figure \ref{fig:co}. All seems to work well, except that one needs an {\it enhancement of a factor 3} to get $P_{2n}$ from theory \cite{Ku90}. Mystery! In {\sl Corradi}'s words: {\it ``that is what happens when theorists do not know what to do"}. Who am I to disagree?
 
{\sl F. Scarlassara} presented $^{60}$Ni+$^{100}$Mo fusion data below the barrier. He claims that the experiment is not conclusive with regard to the possible transfer effect
on deep sub-barrier fusion. {\sl F. Liang} claims that a large sub-barrier fusion enhancement has been observed in reactions with $^{40}$Ca. Comparing to the fusion with $^{48}$Ca, the enhancement in $^{40}$Ca can be attributed to neutron transfer. Multi-nucleon transfer reactions have also been investigated in $^{40}$Ar+$^{208}$Pb, reported by {\sl S. Szilner}. 

{\sl M. Evers} showed data on the reactions $^{208}$Pb($^{16}$O,$^{14}$O)$^{210}$Po, $^{208}$Pb($^{16}$O,$^{12}$O)$^{212}$Po  and $^{208}$Pb($^{16}$O,$^{15}$O)$^{209}$Bi. He also concludes that multi-nucleon transfer processes already play an important role at energies well below the fusion barrier. {\sl C. Beck} also reported on the need of nucleon transfer to explain sub-barrier fusion of $^{32}$S+$^{96}$Zr and $^{40}$Ca+$^{90,96}$Zr. {\sl I. Martel} also discussed effects of neutron transfer on fusion of light halo nuclei at Coulomb barrier energies.

\subsection{Incomplete Fusion, Fission, Evaporation}
 \label{sec:2.7}
 The cross section for fusion evaporation can be written as
 \begin{equation}
 \sigma_{EV}(E)=\sigma_{cap}(E)P_{CN}(E)W_{sur}(E), \label{fe}
 \end{equation}
 where $\sigma_{cap}$ is the capture cross section, $P_{CN}$ is the probability of compound nucleus formation and $W_{sur}$ is the probability of survival through  quasi-fusion processes. Processes such as fusion-fission take the order of $10^{-18}$ s to happen, whereas quasi-fission takes only $10^{-21}$ s. Detectors do not know that, unless they are pretty fast. Humans intervene. The evaporation residue (ER) probabilities are usually calculated by means of the Hauser-Feshbach theory incorporated in popular free numerical codes such as HIVAP, PACE or TALYS.
 
 {\sl P.P. Singh} reported an unexpected increase of incomplete fusion (ICF) for energies above $V_B$. {\sl D.J Hinde} discussed  an improved modeling of ICF.  {\sl A. Wakhle}  presented calculations which show that shell effects around $^{208}$Pb  strongly affect reaction dynamics in ICF reactions. 
 
 {\sl H. Zhang} presented data analysis for the competition between fusion-fission and quasi-fission in the $^{32}$S+$^{184}$W reaction. Figure \ref{fig:zh} shows his interpretation of the measured capture and fusion excitation functions by description of evaporation residue cross sections.
 
{\sl J. Khuyagbaatar} reported measurements of fission cross-sections of $^{34}$S and $^{36}$S induced reactions with $^{204,206,208}$Pb targets.  A larger enhancement of the capture cross-sections below the interaction barriers was observed for $^{34}$S compared to $^{36}$S. 
The experimental results well described except for  $^{204}$Pb (N=122) where the experimental fission cross-sections indicates some enhancement due to higher order channel couplings. A significantly lower experimental ER cross-sections for $^{34}$S compared to calculations show a hint at an additional effect which hinders fusion in the $^{34}$S induced reaction.

{\sl Lukyanov}  discussed the 2n-evaporation channel  in the fusion of $^{4,6}$He+$^{208,206}$Pb reactions, leading to the same compound nucleus.
The excitation functions for the 2n evaporation channels were obtained at energies below the sub-Coulomb barrier region.  A large value of the fusion cross section was observed in the case of the reaction induced by the weakly bound $^6$He projectile due to multi-neutron transfers.

{\sl N. Rowley} claims that the creation of evaporation residues is a very complicated three stage process. The frequently used Eq. \ref{fe} is at best schematic, with $P$ and $W$ ill-defined averages. He proposed a new relation  where all quantities are well-defined, and then he thaught us how to interpret it theoretically. He demonstrated this with respect to the many systems leading to $^{120}$Th for which many good data exist. All this time {\it we have been fooled}, folks! At least now we know it.

{\sl D. Pierroutsakou} presented the results of a systematic study of the excitation of dynamical dipole modes  as a function of the beam energy in fusion reactions leading to the $^{132}$Ce compound nucleus. There is evidence that the prompt dipole radiation is confined at the first moments of the reaction.
{\sl K. Nishio} reported an investigation of fission properties and evaporation residue measurement in reactions using $^{238}$U target nucleus. {\sl K. Masurek} discussed the influence of the potential energy landscape on the fission dynamics.

\begin{figure}
\resizebox{0.9\columnwidth}{!}{%
\includegraphics{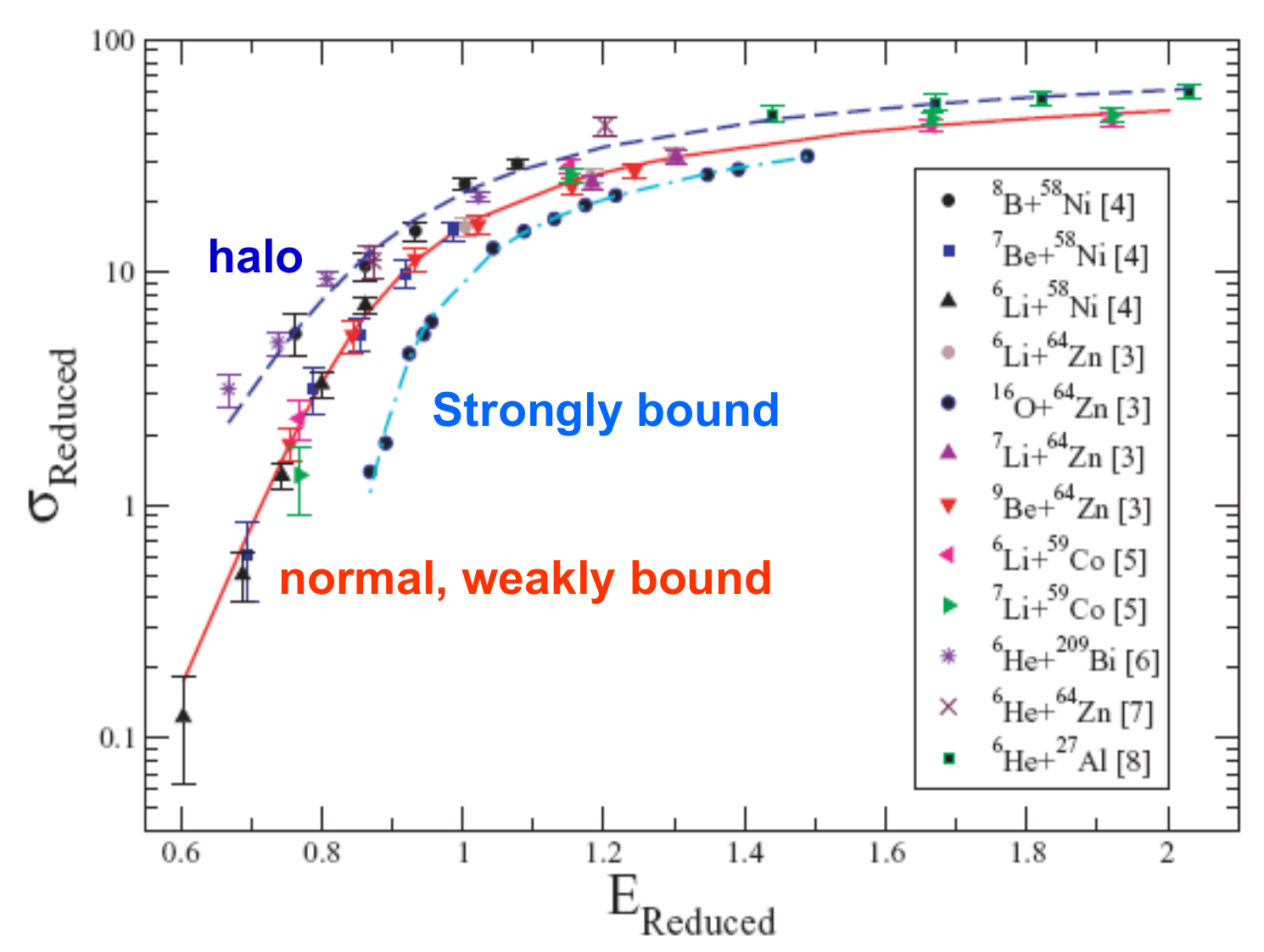}}
\caption{Reduced cross sections for the fusion of halo, normal/weakly bound, and strongly bound nuclei.  (Courtesy of {\sl J. Kolata}).
}
\label{fig:ko}       
\end{figure}

{\sl D. Mancusi} discussed the
constraining of statistical-model parameters using fusion and spallation reactions. {\sl G. Ademard} reported studies on the decay of excited nuclei produced in the $^{78,82}$Kr+$^{40}$Ca reactions at 5.5 MeV/nucleon. 

\section{Special Topics}
\label{sec:2.8}
\subsection{Rare Isotopes}
 \label{sec:2.9}
{\sl P. Gomes} introduced a Universal Fusion Function (UFF) to investigate the role of breakup dynamical effects on fusion of neutron halo $^6$He weakly bound systems. The idea is that one divides and multiplies the calculated fusion cross sections by factors which account for the major physics behind fusion. Then the deviations from the expected UFF are a hint of coupled channel effects (unknown physics, I told you.). By comparing the UFF with data on fusion of $^6$He and $^{11}$Be on numerous targets, he concludes that there is a suppression above the barrier and an enhancement below the barrier. The UFF idea reminds me of efforts to find a {\it grand unified theory of everything} in particle physics. Ambitious, specially for nuclear physicists.

\begin{figure}
\resizebox{0.8\columnwidth}{!}{%
\includegraphics{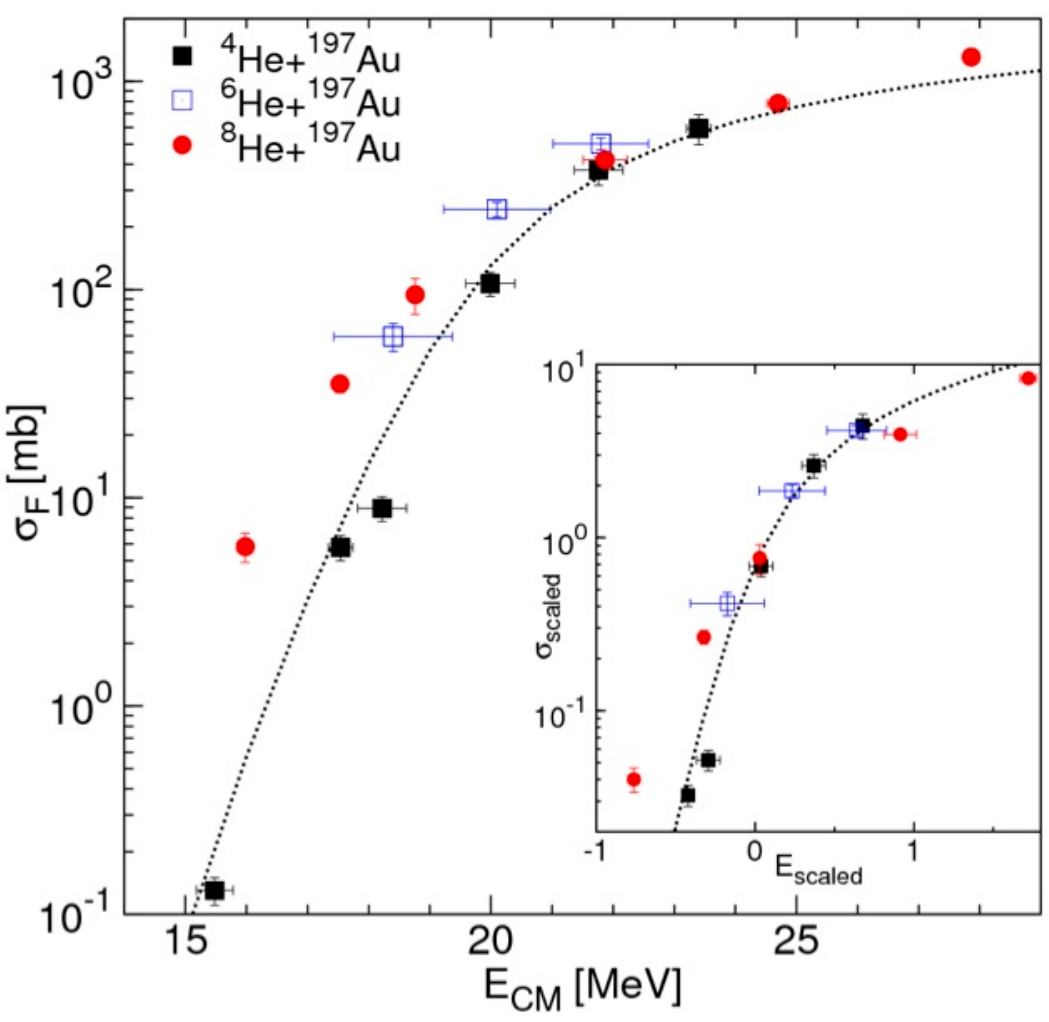}}
\caption{Fusion cross sections of $^{4,6,8}$He with $^{197}$Au.The inset is the reduced (scaled) cross sections.  (Courtesy of {\sl A. Lemasson}) 
}
\label{fig:lem}       
\end{figure}

{\sl M. Alcorta} and {\sl E. Rehm} presented new measurements of fusion-fission cross sections for the systems $^{13,14,15}$C+$^{232}$Th. They conclude that fusion of
$^{14}$C+$^{232}$Th is similar to that of $^{12,13}$C+$^{232}$Th and that fusion of $^{15}$C+$^{232}$Th shows a fusion enhancement by a factor of 5-6 at $E/V_B\sim 0.85$. {\sl A. Di Pietro} showed a damping of elastic cross-section for the reaction induced by the $^{11}$Be nucleus when compared with both $^9$Be ($S_n=1.67$ MeV) and $^{10}$Be ($S_n=6.8$ MeV). 

{\sl E. F. Aguilera} and {\sl J. Kolata} presented new data on evaporation protons from $^8$B+$^{58}$Ni at 8 energies. Perhaps the most impressive result from this group is shown in figure \ref{fig:ko}. One sees the reduced (scaled by geometry) cross sections  for the fusion of halo, normal/weakly bound, and strongly bound nuclei.  A clear tendency is seen of fusion enhancement with decreasing binding energy. 

{\sl V. Guimar\~aes} discussed an optical model  analyses with double-folding S\~ao Paulo Potential for $^{7,8,9}$Be+$^{12}$C elastic scattering. {\sl M. Mazzocco} presented data on quasi-elastic angular distributions for $^{17}$F. The collected data were analyzed within the framework of the optical model with the coupled-channels to extract the reaction cross sections and to investigate the relevance of direct reaction mechanisms. Only a very small influence of the $^{17}$F low binding energy on the reaction dynamics was found. 

{\sl W. Loveland} claims that $^9$Li fusion excitation functions show sub-barrier fusion enhancement which are not easily accounted for by current models of fusion. He also mentioned that to understand this better one would need to measure the fusion of $^{11}$Li+$^{208}$Pb, the {\it ``holy grail"} of fusion reactions. This reminds me of my preferred    Monty Python quote: {\it ``we are the knights who say  ni}". 

{\sl G. Potel} presented a calculation of nucleon-pair transfer in  reactions with neutron-rich nuclei using concepts of the BCS theory. He finds that the absolute cross section associated with the first excited state of $^9$Li in the p($^{11}$Li,$^9$Li*($1/2^-$; 2:69MeV))t reaction is a {\it direct evidence of phonon mediated pairing in nuclei}. His method is an extension of Eqs. \ref{transf} and \ref{transf2}, with $P_{\beta\alpha}$ now including four nucleon wavefunctions for simultaneous two-neutron transfer. He also includes sequential two-neutron transfer. In this case a nucleon propagator for the intermediate state is included.  

{\sl L.V. Grigorenko} discussed new theoretical advances in studies of two-proton radioactivity and three-body decays. The lifetime and decay energy systematics for several known and prospective true 2p emitters were calculated.

{\sl Sh. A. Kalandarov} reported the production of the doubly magic nucleus $^{100}$Sn in $^{72,74,76}$Kr+$^{40}$Ca, $^{72,74,76}$Kr+$^{40}$Ar and $^{72,74,76}$Kr+$^{32}$S reactions at 4-6 MeV/nucleon. {\sl N. Madhavan} reported a new gas-filled separator to be built at New Delhi. The relevance of breakup of $^{6,7}$Li in fusion reactions was discussed by {\sl D. Luong}. {\sl A. Drouart} reported a new toy: the $S^3$ - Super Separator Spectrometer - at GANIL.

{\sl A. Lemasson} showed new data on fusion of $^{4,6,8}$He. He reported an unexpected similar behavior of the cross sections for $^6$He and $^8$He \cite{Lem09}: the additional two neutrons do not modify the tunneling probability (see figure \ref{fig:lem}). These results might be a showcase for the general problem of {\it tunneling of composite objects} \cite{BFZ07}: in loosely-bound systems, tunneling of clusters might occur with different time-scales. Similar study was presented by {\sl Y. Penionzhkevich}.

\subsection{Clusters}
 \label{sec:2.10}
 
\begin{figure}
\resizebox{1.0\columnwidth}{!}{%
\includegraphics{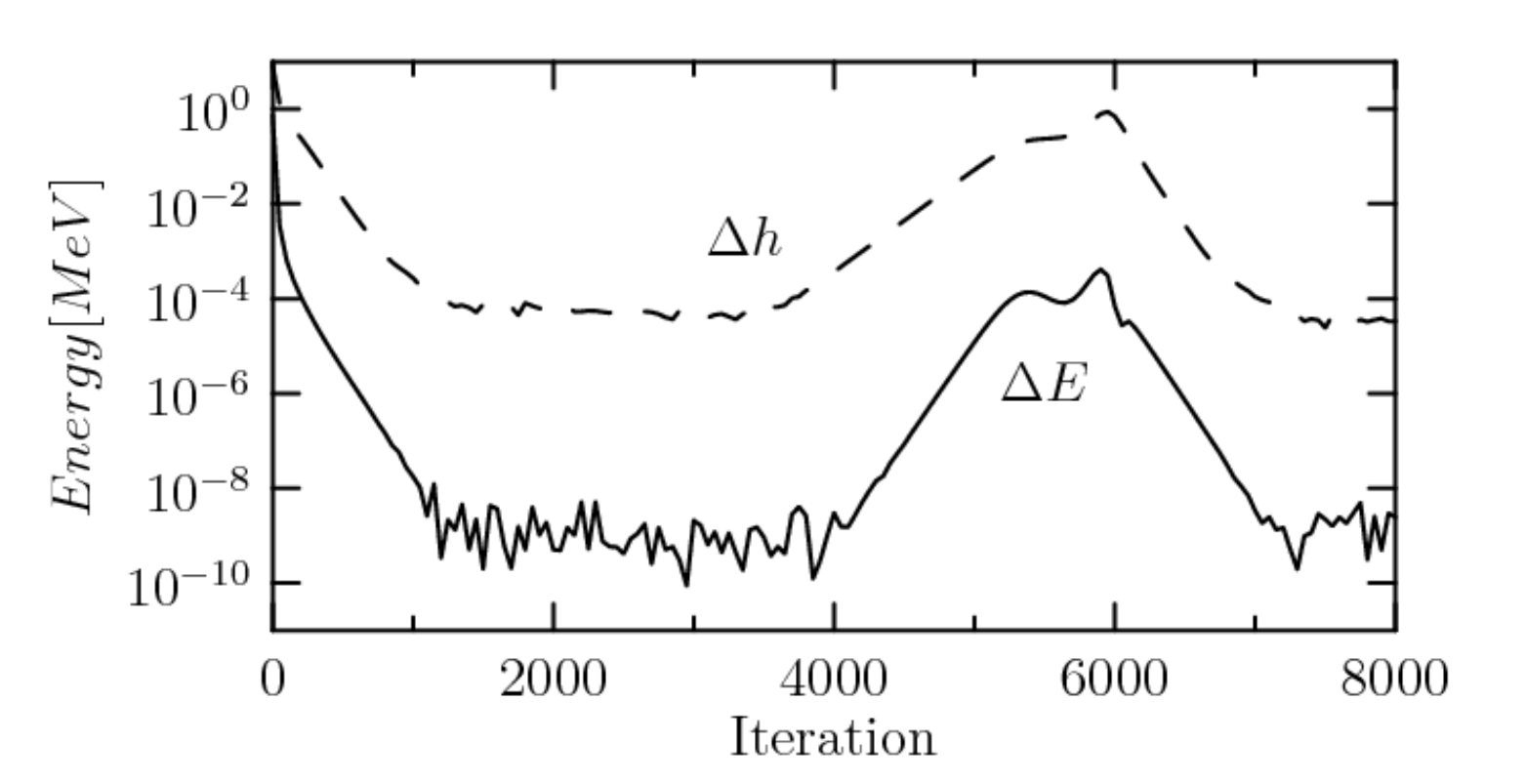}}
\caption{The discovery of ``numerical tunneling". (Courtesy of {\sl J. Maruhn}).
}
\label{fig:ma}       
\end{figure}

 {\sl J. Maruhn} discussed the new advances in $\alpha$-cluster formation in nuclei based on a time-dependent Hartree-Fock (TDHF) method. Results showing $\alpha$-chain states in $^{12,16,20}$C were presented. Also, a  chain state in $^{16}$O  provides  rotational stabilization. {\sl Maruhn} also announced the discovery of {\it ``numerical tunneling"}. This is shown in figure \ref{fig:ma}. A convergence indicator is used to assess if  the ground state of the hamiltonian $h$ is reached, $$\Delta h={1\over A} \sum_k\sqrt{\langle \phi_k|\hat{h}^2|\phi_k\rangle -\langle \phi_k|\hat{h}|\phi_k\rangle^2}.$$
An excited quasi-stable state appears as an apparently converged configuration for 1000's of iterations. But then, suddenly, unexpectedly, and also puzzlingly, there is convergence path to the ground state via triaxial shapes. This only happens if you allow the computer code to run way beyond what you (wisely) would ask it to do. Isn't that worth an international prize? At least the {\it Ig nobel prize}, please.

{\sl M. Ito} reported studies of light neutron-excess systems from bound to continuum states. His method is based on molecular states in nuclei. In particular, he calculated highly excited states of Be isotopes and in $^x$He+$^y$He reactions. He observed cluster structures in light 4N nuclei and in $\alpha$-cluster nuclei such as $^8$Be=2$\alpha$, $^{12}$C=3$\alpha$, $^{16}$O=$\alpha$+$^{12}$C. He also showed the monopole response of $^{12}$Be and its ratio to single-particle response in the continuum.

\subsection{Nuclear Astrophysics}
 \label{sec:2.11}

{\sl A. Guglielmetti}  presented recent results on (p,$\gamma$) and ($\alpha,\gamma$) fusion reactions at the LUNA facility. In particular, she presented the new S-factor for $^3$He+$^4$He reaction, $S_{34}=0.567\pm0.018\pm0.004$ keV b. The reduced uncertainty due to $S_{34}$ on neutrino flux implies a reduction $\Phi_{^8B} =7.5\% \rightarrow 4.3\%$ and  $\Phi_{^7Be} =8\% \rightarrow 4.5\%$.

{\sl F. de Oliveira Santos} presented new results for $^{18}$F(p,$\alpha$)$^{15}$O, H($^{17}$Ne,p)$^{17}$Ne, H($^{14}$O,p)$^{14}$O. {\sl R.G. Pizzone} discussed the ``{\it Trojan Horse Method}"  (THM) used to determine the cross section for the $^6$Li(d,$\alpha$)$^4$He reaction. The excitation function obtained from direct data of the $^6$Li(d,$\alpha$)$^4$He and $^7$Li(p,$\alpha$)$^4$He is well reproduced by THM below and above the Coulomb barrier in both cases, which attests theTHM particle invariance, or {\it pole invariance}.

{\sl B. Jurado}  presented new results on surrogate reactions, e.g., (n,f) obtained from transfer reactions. A major challenge for this method is to show that the population of states via a ``{\it surrogate reaction}" is the same as that for capture of a free neutron. In the case of independence on the population of angular momentum states $J^\pi$, the reaction can be described by the Ewing-Weisskopf theory (EWT). Hauser-Feshbach theory is  the same as EWT with proper account of angular momentum. A successful case is shown in figure \ref{fig:ju}. Excellent agreement of EWT calculations at low energies for this case, shows that  the fission cross sections are not sensitive to differences $J^\pi$ distributions \cite{Ke10}. Unfortunately, this seems to be more an exception that the rule. 
 
{\sl A. Goasduff} reported measurements of $^{12}$C+$^{16}$O sub-barrier radiative capture cross sections.
{\sl K. Czerski} and {\sl J. Kasagi} presented results on enhanced electron screening in nuclear reactions
Their data show a much enhanced fusion cross section within liquids and metals, which cannot be explained by theory \cite{Cz01}. This unavoidably reminds me of the {\it cold fusion saga}. Let us hope that it really works. In this way we could avoid paying the bill for ITER and other machines. It would also save a lot of space in France.

\begin{figure}
\resizebox{1.0\columnwidth}{!}{%
\includegraphics{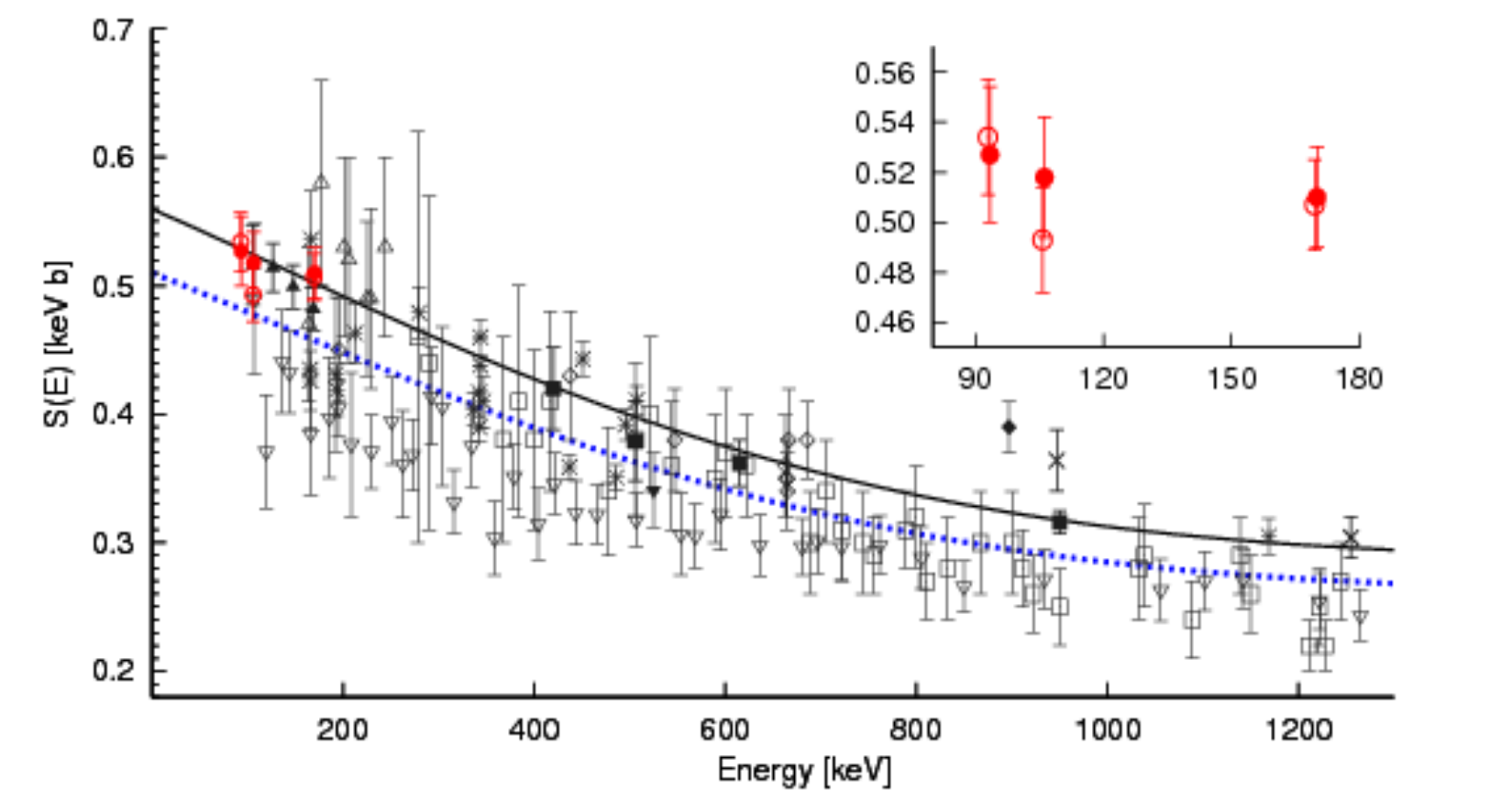}}
\caption{New S-factor for $^3$He+$^4$He reaction from the LUNA collaboration.  (Courtesy of {\sl A. Guglielmetti}).
}
\label{fig:le}       
\end{figure}

\subsection{Superheavy Elements}
 \label{sec:2.12}

In the 1960's a famous Russian professor was invited for  a conference in France. ``Professor, this was very nice", said the conference chairman at the end of his talk, ``but I did not understand why you talked for 5 minutes and sat down for another 5 minutes, repeatedly during your seminar", continued the chairman. ``I am very sorry", replied the Professor, ``but I was told by my comrades in Russia that the French are very slow thinkers. Thus I was giving you time to understand what I said".  If you participated in this conference, you know that I am not joking.

{\sl J. Hamilton}  reported on the discovery of the new superheavy element (SHE) with Z=117. This was a mark achievement in this field, and was only possible with the finding that one needed $^{249}$Bk to produce the fusion of $^{48}$Ca and $^{249}$Bk, followed by 3n evaporation \cite{Og10}. Berkelium was produced and bought from the Oak Ridge National Lab at a very salty price and shipped to Dubna/Russia where the superheavy science was done. It is a clear realization that the US has turned from a major science achiever to an {\it exporter of raw material}, a real B-republic (B is for Berkelium, not Banana).

{\sl K. Morita} presented the latest efforts at RIKEN on (a) the reaction $^{209}$Bi($^{70}$Zn,n)$^{278}$113, with a cross-section of $18+25-13$ fb,  (b) new spectroscopic data on $^{266}$Bh, $^{262}$Db  with further confirmation of $^{278}$113,  and $^{265}$Sg$^{a/b}$, $^{261}$Rf$^{a/b}$ with further confirmation of $^{277}$Cn, and on $^{264}$Hs, $^{263}$Hs. A dedicated community work.

{\sl A. Karpov} discussed ternary quasi-fission of giant nuclear systems. True ternary fission is impossible for actinides (insufficient mass). Superheavy nuclei have a real chance to split onto tin + something + tin. Giant nuclear molecules may decay onto lead + something + lead.

\begin{figure}
\begin{center}
\resizebox{0.8\columnwidth}{!}{%
\includegraphics{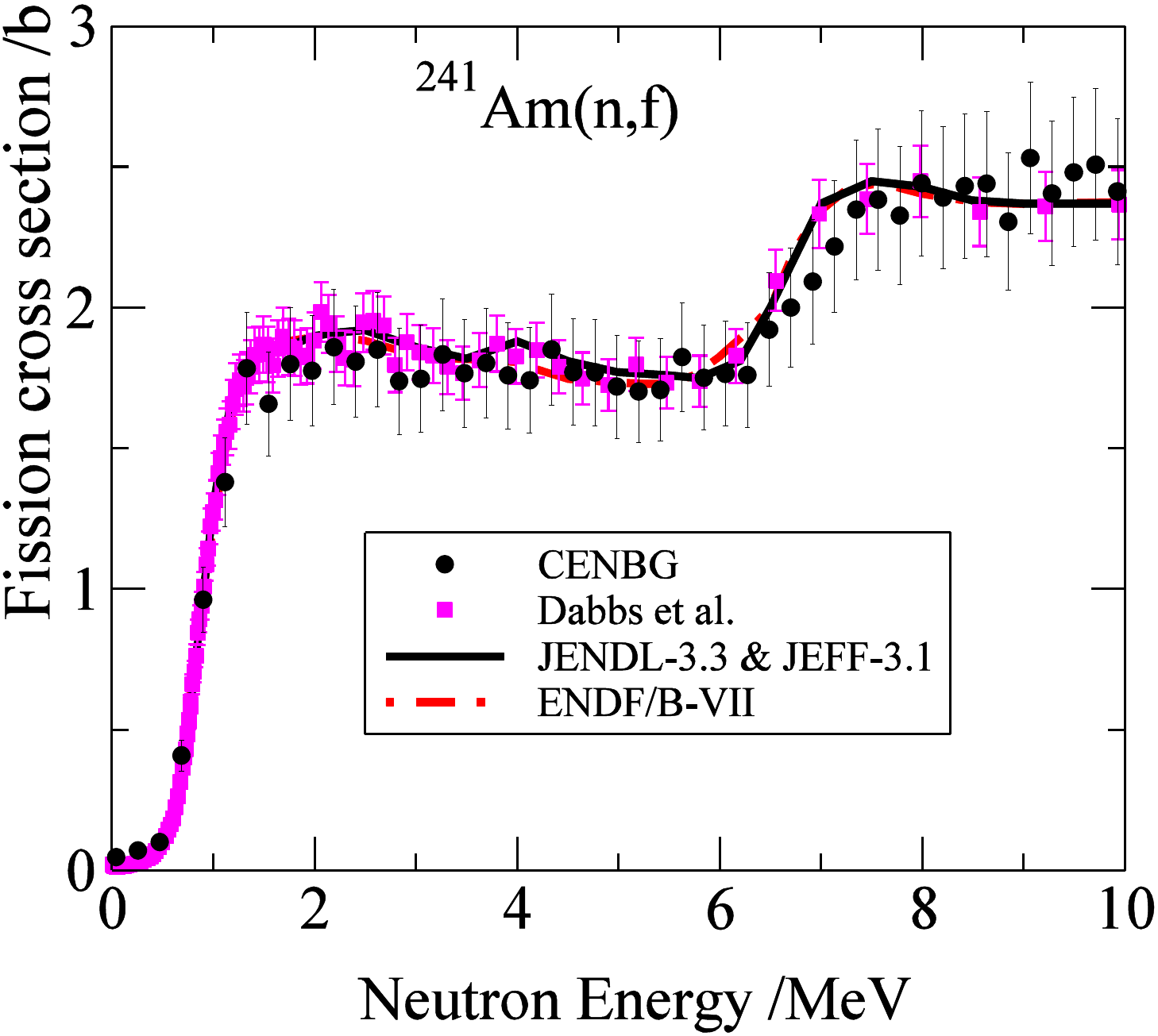}}
\end{center}
\caption{Fission cross sections following neutron capture on $^{241}$Am obtained from the surrogate method. The data is well explained without need for accounting for  $J^\pi$ distributions.  (Courtesy of {\sl B. Jurado}).
}
\label{fig:ju}       
\end{figure}

{\sl A. A. Voinov} discussed the reaction $^{226}$Ra+$^{48}$Ca=$^{269}$-$^{271}$Hs+3-5n. Six decay chains of $^{270}$Hs were observed at 233 MeV beam energy. A cross section $\sigma_{4n}= 8.3$ pb was measured to be lower than predicted. No decay chains of $^{269-271}$Hs isotopes were observed at two other bombarding energies of 228.5 MeV and 240.5 MeV. The upper cross section limits are $\sigma_{3n} < 4.2$ pb and $\sigma_{5n} < 5.0$ pb for the low and high $^{48}$Ca beam energy, respectively

\begin{figure}[t]
\resizebox{1.0\columnwidth}{!}{%
\includegraphics{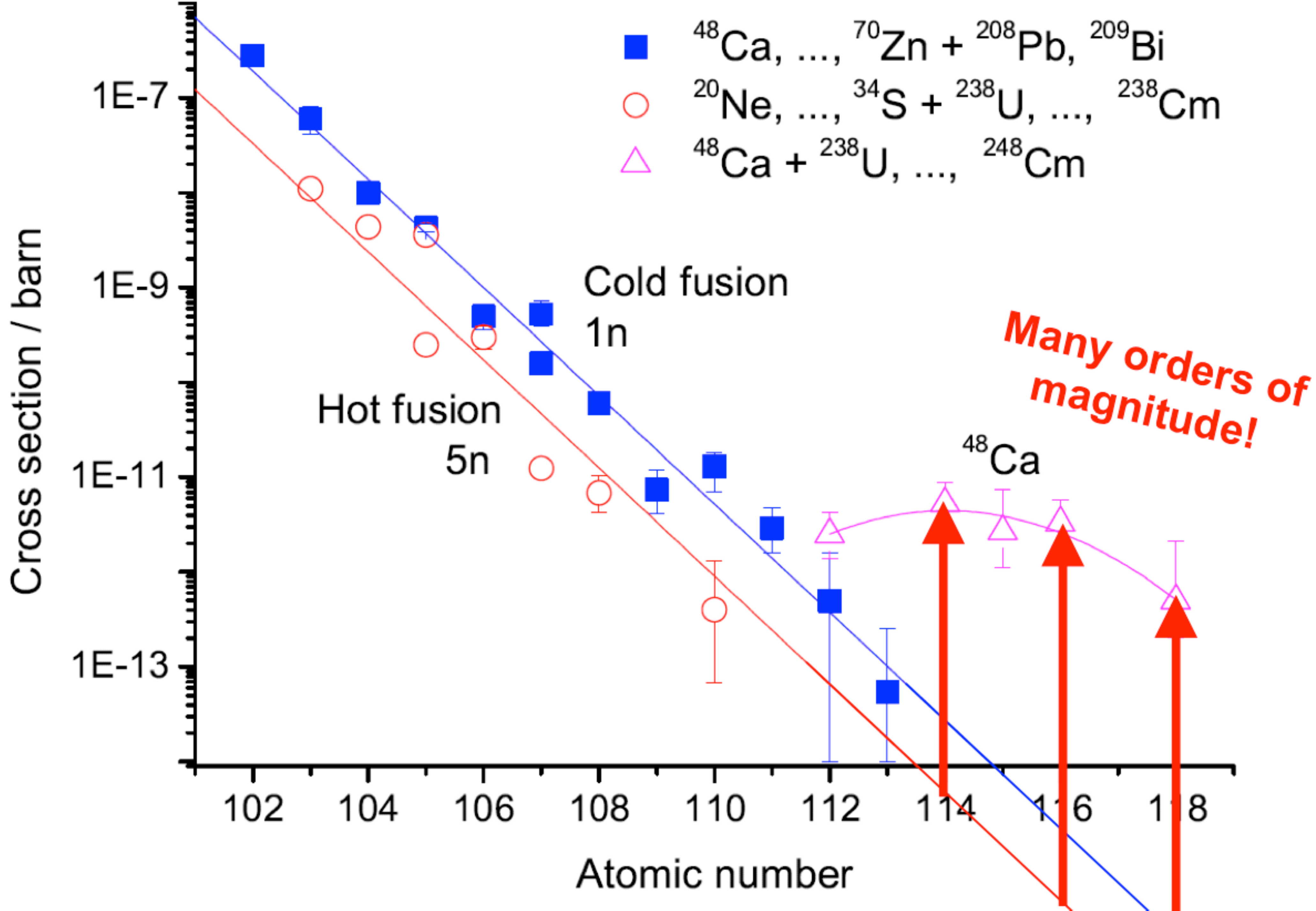}}
\caption{Fusion cross sections leading to the formation of superheavy elements.  (Courtesy of {\sl Christoph E. D\"ullmann}).
}
\label{fig:ju}       
\end{figure}

{\sl M. Itkis} taught us how to measure reactions with $^{48}$Ca. $^{48}$Ca is a {\it best kept secret} by the Russians. With $^{48}$Ca the newest elements were found. Why did it take so long to realize that $^{48}$Ca was THE nucleus? The double-magic nucleus $^{48}$Ca allows one to obtain the low excitation energy of compound nucleus ($E^*\sim 30-36$ MeV) at the Coulomb barrier. The neutron excess leads to $N_{CN}=170-180$ in the reaction with actinide targets in contrast to cold fusion reactions, where $N_{CN}\sim150-160$. 
The heaviest element, which can be obtained with the reactions with $^{48}$Ca-projectiles, is the 118 nucleus. {\sl Itkis} claims that a possible alternative pathway for SHE synthesis is represented by the complete fusion of actinide nuclei with heavier projectiles such as $^{58}$Fe or $^{64}$Ni leading to the formation of CN with $Z=118-124$ and $N=178-188$.

{\sl P. Armsbruster} claims that it is hard to understand the production cross sections for reactions induced by beams beyond $^{48}$Ca. In fact, it seems to be hard to understand the unexpectedly high cross sections with $^{48}$Ca. One needs to carry out experiments to determine the atomic numbers of the elements $Z=114-118$, either by chemistry or by characteristic $K$ and $L$ $X$-ray energies.
{\sl Armbruster} also mentioned that fission of oblate nuclei has never been observed. Their fission probabilities should be measured. By the way, {\it Happy 80$^{th}$ Birthday, Peter!} We hope that you continue to help us understand the nature of fusion and fission for a long time.

{\sl Christoph E. D\"ullmann} discussed  the TASCA research program where element 114 was identified with high cross sections (10 pb) and high efficiency (60\%), what open up new avenues for other experiments. Next experiments will focus on direct Z determination of $^{48}$Ca+$^{243}$Am products and the search for element 120. 

Finally, {\sl V. Zagrebaev} taught us that the time-dependent Schr\"odinger equation shows us (only if you ask it right) that at low-energy collisions nucleons do not ``jump" from one nucleus to another. The wave functions of valence nucleons follow the
{\it two-center molecular states} spreading over both nuclei. Two-Center Shell Model + Adiabatic Potential Energy Surface + Transport (Langevin type) Equations of Motion are appropriate for description of low-energy multi-nucleon transfer \cite{Za03}. In the end, he presented new ideas towards heavier elements:  (a) produce SHE with pulsed nuclear reactors, (b) produce SHE in multiple (rather soft!) {\it nuclear explosions} \cite{Bot10}. Wow! I like the ideas,  even the last one. But I only hope that they perform this experiment within the borders of the Russian Federation.

The meeting was a huge success because of the dedicated work of the organizers   Navin Alahari (chair), H\'eloise Goutte, Denis Lacroix, Christine Lemaitre, Maurycy Rejmund and Christelle Schmitt. We are thankful for their community service and for the usual great French hospitality (and for the delicious food, of course). 

The next FUSION conference was chosen by the International Advisory Committee to be held in {\it New Delhi, India, 2014}. I can't wait.
  
 \medskip
 
Work supported by the US DOE grants FG02-08ER41533, SC0004971, and FC02-07-ER41457 (UNEDF).


\begin{thebibliography}{}
\bibitem{SPP} L. C. Chamon, et al., Phys.
Rev. Lett. 79 (1997) 5218.
\bibitem{Na85} M.A. Nagarajan et al., Phys. Rev. Lett. 54 (1985) 1136.
\bibitem{Hu06} M.S. Hussein et al., Phys. Rev. C 73 (2006) 044610.
\bibitem{Ji04} C.L. Jiang et al., Phys. Rev. Lett. 93 (2004) 012701.
\bibitem{Es78} H. Esbensen et al., Phys. Rev. Lett. 41 (1978) 296.
\bibitem{RS91} N. Rowley, G.R. Satchler and P.H. Stelson, Phys. Lett. B 254 (1991) 25.
\bibitem{Tim95} H. Timmers et al., Nuc. Phys. A584 (1995) 190.
\bibitem{EB07} H. Esbensen and S. Misicu, Phys. Rev. C 76 (2007) 054609.
\bibitem{IHI07} T. Ichikawa, K. Hagino, and A. Iwamoto,  Phys. Rev. C 75 (2007) 057603.
\bibitem{Ku90} R. K\"unkel et al.,  Z. Phys. A 336 (1990) 71.
\bibitem{SA10} V.V. Sargsyan, G.G. Adamian, N.V. Antonenko, and W. Scheid, Eur. Phys. J. A 45 (2010) 125.
\bibitem{Lem09} A. Lemasson et al., Phys. Rev. Lett. 103 (2009) 032701.
\bibitem{BFZ07} C.A. Bertulani, V. Flambaum and V.G. Zelevinsky, J. Phys. G 34 (2007) 2289.
\bibitem{Ke10} G. Kessedjian et al., Phys. Lett B 692 (2010) 297.
\bibitem{Cz01} K. Czerski et al., Europhys. Lett., 54  (2001) 449.
\bibitem{Og10} Yu. Ts. Oganessian et al., Phys. Rev. Lett. 104 (2010) 142502.
\bibitem{Za03} V.I. Zagrebaev, Phys. Rev. C 67 (2003) 061601.
\bibitem{Bot10} A. Botvina et al., arXiv:1006.4738.
\end{thebibliography}
\end{document}